\documentclass[useAMS,usenatbib,usegraphicx]{mn2e}

\usepackage{amssymb}                                            %%
\usepackage{amsmath} 
\usepackage{array,tabularx}
\newcommand{\diff}[2]{\ensuremath{\frac{\mathrm{d}\,{#1}}{\mathrm{d}\,{#2}}}}
\def\aap{A\&A}%
\def\apjl{ApJ}%
\def\apj{ApJ}%
\def\mnras{MNRAS}%
\def\apss{Ap\&SS}%
\def\araa{ARA\&A}%
\def\prd{Phys.~Rev.~D}%
\def\nat{Nature}%
\title[Population synthesis  of gamma-ray bursts and the spinar paradigm]{Population synthesis  of gamma-ray bursts with precursor
activity and the spinar paradigm}
\author[G. V. Lipunova et al.]
{G. V. Lipunova$^{1}$\thanks{E-mail: galja@sai.msu.ru}, 
E. S. Gorbovskoy$^{1,2}$\footnotemark[1]\thanks{E-mail: gorbovskoy@sai.msu.ru},
A. I. Bogomazov$^{1}$,
V. M. Lipunov$^{1,2}$\\
$^{1}$Sternberg Astronomical Institute, Universitetskiy pr., 13, Moscow 
119992, Russia\\
$^{2}$Moscow State University, Vorobievy Gory, Moscow 119991, Russia
}
\begin{document}
\date{Accepted 2009 May 16. Received 2009 May 07; in original form 2009 March 18;\newline 
The definitive version is available at www.blackwell-synergy.com\newline
http://www3.interscience.wiley.com/cgi-bin/fulltext/122513307/HTMLSTART\newline
doi:10.1111/j.1365-2966.2009.15079.x}

%\firstpage{1695}
%\lastpage{1704}
\pagerange{\pageref{firstpage}--\pageref{lastpage}} 
\pubyear{2009}
\volume{397}
\pagerange{1695--1704}
\maketitle

\label{firstpage}

\begin{abstract}
We study statistical properties of long gamma-ray bursts (GRBs) produced
by the collapsing cores of WR stars in binary systems. Fast rotation of 
the cores enables a two-stage collapse scenario, implying the formation
of a spinar-like object. A burst produced by such a collapse consists of
two pulses, whose energy budget is enough to explain observed GRBs. We
calculate models of spinar evolution using results from a population
synthesis of binary systems (done by the `Scenario Machine') as initial
parameters for the rotating massive cores. Among the resulting bursts,
events with the weaker first peak, namely, precursor, are identified,
and the precursor-to-main-pulse time separations fully agree with the
range of the observed values. The calculated fraction of long GRBs with
precursor (about 10 per cent of the total number of long GRBs) and the
durations of the main pulses are also consistent with observations.
Precursors with lead times greater by up to one order of magnitude than
those observed so far are expected to be about twice less numerous.
Independently of a GRB model assumed, we predict the existence of
precursors that arrive up to $\gtrsim 10^3$~s in advance of the main
events of GRBs.
\end{abstract}


\section{Introduction}
\label{s:intro}

Gravitational collapse is believed to be the underlying mechanism for
the most energetic events observed in the Universe: GRBs and supernovae.
While it is commonly accepted that the remnant of such events is a black
hole or a neutron star, the details of the process are uncertain. In
relation to GRBs, we investigate here the collapse of a fast rotating 
magnetized object, which can be understood in terms of the `spinar
paradigm'. We define spinar as a critically-fast rotating magnetized
relativistic object, whose quasi-equilibrium is maintained by the
balance of centrifugal and gravitational forces. The evolution of a
spinar is determined by its magnetic field. A benefit of the spinar
model is that it describes transparently and in a simple way the main
features of a real collapse.

The properties of rotating magnetized objects were first investigated to
understand the mechanisms of active galactic nuclei by, e.g.,
\citet{hoyle-fowler1963,ozernoy1966e, morrison1969,
woltjer1971,bisnov-blinnikov1972e,ozernoy-usov1973}, of pulsars by
\citet{gunn-ostriker1969a}, and of supernova explosions by
\citet{leblanc-wilson1970, bisnovatyi-kogan1971e}. A rotation-supported
`cold' configuration with magnetic field received the name
`spinar'~\citep[see early reviews by][]{morrison-cavaliere1971,
ginzburg-ozernoi1977e}. Stellar mass spinars were suggested
by~\citet{lipunov1983,lipunov1987}. In the works by
\citet{lipunova1997e} and \citet{lipunova-lipunov1998} a burst of
electromagnetic radiation produced during the collapse of a spinar was
studied, and a spinar mechanism for GRBs was first suggested. 

As \citet{lipunov-gorb2007} point out, there should be energy release in
a process of spinar formation as well. This approach enables one to
consider a two-stage scenario of a collapse. At the first stage, a
spinar forms from a collapsing rotating body when centrifugal forces
halt contraction. The effective dimensionless Kerr parameter of the
spinar is greater than unity. At the second stage, the angular momentum
is carried away, and the spinar evolves to a limiting Kerr black hole or
a neutron star, depending on its mass.

\citet{lipunov-gorb2008} develop a 1D model of the magneto-rotational
collapse of a spinar, which includes all principle relativistic effects
on the dynamics and the magnetic field, along with the pressure of
nuclear matter and neutrino cooling. A variety of burst patterns is
obtained, generally a combination of two peaks. It is shown that the
spinar paradigm agrees with the basic observed GRB properties.

Potential progenitors of spinars are rotating WR stars without H and He
envelops, which are already considered as possible progenitors of GRBs
~\citep[for a review, see][]{woosley-bloom2006}. The spinar mechanism
requires the presence of a high angular momentum in the WR core at the
start of the collapse. A direct collapse to a black hole is impossible
if the rotating core has an effective dimensionless Kerr parameter
greater than unity:
\begin{equation}
\label{eq:kerr} a\equiv\frac{I\,\Omega}{G\,M_\mathrm{c}^{\,2}/c} > 1,
\end{equation}
where $I$ is the moment of inertia of the core, $\Omega$ is the angular
velocity, $M_\mathrm{c}$ is the core mass.  Equation~\eqref{eq:kerr}
corresponds to the following condition  on the specific angular
momentum:
$$
\frac{I\,\Omega}{M_\mathrm{c}} > 4.4\times 10^{15}
\frac{M_\mathrm{c}}{M_\odot}\, \mbox{cm}^2\,
\mbox{s}^{-1}\, .
$$
Fast rotation can be a result
of the evolution of a rotating single massive star or a star in a close
binary system \citep[see, e.g.,][]{vandenHeuvel-yoon2007}.

Different scenarios for the collapse of a WR core are possible depending
on the unknown properties of the collapsing core, among them the
quantity and distribution of the angular momentum within the core. These
scenarios have so far included: a highly magnetized rotating neutron
star, a black hole surrounded by an accretion disc (the `collapsar'
model), a hypermassive rotating neutron star with an accretion
disc~\citep[see references in][]{woosley-bloom2006}. 

In this regard, the spinar mechanism is a natural complement to the
range of potential GRB producers. It is important to note that in
distinction from the collapsar model of ~\citet{woosley1993}, in the
spinar paradigm a GRB begins with a spinar formation and continues with
its collapse to a limiting black hole. In the collapsar model, we first
have the formation of a black hole, and after that a GRB develops,
powered from an accreting disc-like envelope and by the Blandford-Znajek
mechanism. It is essential for the spinar that the central part of a
core with mass $\sim 2-3 M_\odot$ has large angular momentum (effective
Kerr parameter $>1$). In contrast, the collapsar model requires that
there is less angular momentum in the centre and an excess at the
periphery.

There are calculations of the structure of WR cores supporting a
hypothesis that may be there is too much angular momentum in the centre.
For example, the results of \citet{hirschi_et2005,yoon_et2006} indicate
that the inner part of a WR core with mass $\sim 2.5~M_\odot$ is
characterized by the effective dimensionless Kerr parameter not less (or
not significantly less) than an effective Kerr parameter of the whole
core. Consequently, if a core has the effective Kerr parameter
considerably greater than unity and cannot undergo a direct collapse to
a black hole, then the inner part is equally prohibited to do so. The
results of such models, usually intended for single stars, are generally
uncertain as they are strongly dependent on the physics assumed. 

In the present work we consider only massive stars collapsing in binary
systems. In a binary system, tidal interaction and synchronization lead
to the fast rotation of a massive pre-collapse
core~\citep{tutukov-yungelson1973r}. Binary systems as GRB sites were
considered by, e.g., ~\citet{woosley1993,
paczynski1998,brown_et2000,postnov-cherep2001,
tutukov-cherep2003e,tutukov-cherep2004e,
izzard_et2004,podsiadlowski_et2004,
petrovic_et2005,bogomazov_et2007,vandenHeuvel-yoon2007}.

 In previous work~\citep{lipunov-gorb2008}, the dependence of
qualitative features of GRBs on the initial parameters of collapsing
cores (their angular momentum, magnetic flux, and mass) have been
studied by calculating numerous models on a uniform grid of parameter
values. It is clear, however, that the initial parameters are not
distributed uniformly, and their values are defined by the course of
previous evolution.

To derive the initial distribution on the effective Ker parameter, 
we use a population synthesis
of binary stars carried out by the Scenario
Machine~\citep{lipunov_et1996a, lipunov_et2007}. We perform population
synthesis for different binary evolution parameters and analyse the
resulting set of GRBs, which can be classified into different kinds of
events.

Below we focus particularly on the precursor phenomenon~\citep[][and
references
therein]{koshut_et1995,lazzati2005,burlon_et2008,wang-meszaros2007},
regarding it as a primary energy release accompanying a spinar
formation. The `lead time' of a precursor, namely, the separation time
from the main pulse, is an important and robust parameter that we
calculate. As we show, there should be GRBs with precursor occuring
hunderds and thousands of seconds in advance. We also investigate the
dependences between energy and temporal characteristics of two GRB
pulses.

In the next section we give a brief account of the relevant aspects of
the population synthesis of binary stars. In
Sect.~\ref{s:spinar_calculation} we describe calculation of spinar
evolution. Results on GRB statistics are presented in
Sect.~\ref{s:results}. Discussion follows in Sect.~\ref{s:discussion}.

\section{Population synthesis of WR stars in binary systems}
\label{s:popsyn}

Population synthesis of binary systems is performed by the `Scenario
Machine', described in detail by \citet{lipunov_et1996a,
lipunov_et2007}. Population synthesis of WR stars in binary systems as
progenitors of long GRBs is done by
\citet{bogomazov_et2007,bogomazov_et2008e} to investigate the rates and 
correlation with the host-galaxy morphology. Here we describe only a few
elements of the computer code, which involve the parameters varied in
the present study.

The minimum initial mass of a star that may eventually produce a black
hole is set to $25 M_\odot$~\citep{tutukov-cherep2003e}.\footnote{If in
the course of the evolution a star accretes matter from the companion,
we check the maximum mass it attains after the accretion.} Its companion
can have any mass between $0.1\, M_\odot$ and the mass of the main star,
so the mass ratio $q=M_2/M_1<1$. The initial masses of the main stars
are distributed by Salpeter's law above $25 M_\odot$. As representative
cases, two initial distributions of mass ratio $f(q)=q^{\alpha_q}$ are
considered: with ${\alpha_q}=0$ and ${\alpha_q}=2$. 

Two models of a stellar wind are used:

\begin{enumerate}
\renewcommand{\theenumi}{W\arabic{enumi}:}

\item `Weak stellar wind'. This scenario corresponds to that named `A'
in \citet{bogomazov_et2007, lipunov_et2007}. On the main sequence and at
the supergiant stage, the total mass loss does not exceed 20 per cent of
the initial mass. During the WR-star stage, the mass loss is 30 per cent
of the whole star mass. Weak stellar winds are expected for stars with
low metallicity.

\item `Strong stellar wind'. This scenario corresponds to that named `C'
in \citet{bogomazov_et2007, lipunov_et2007}, except that a star loses 
70 per cent of its envelope at each evolutionary stage, not the whole
envelope. Note that, in the original model `C', GRBs cannot be produced
because the strong wind results in a too large semi-major axis of the
orbit before the collapse of a binary
component~\citep{bogomazov_et2007}. 

\end{enumerate}

In model W1, the mass loss rate $\dot M$ for a main sequence star is
described by the formula $ \dot M=\alpha_\mathrm{w}\,
L\,c/V_{\infty}\,$, where $L$ is the star luminosity, $V_{\infty}$ is
the wind velocity at infinity, $c$ is the speed of light,
$\alpha_\mathrm{w}$ is a free parameter. On the main sequence and at the
supergiant stage, the mass variation in model W1 during any stage does
not exceed a value of $0.1\,(M-M_\mathrm{core})$, where $M$ is the mass
of the star at the beginning of the stage and $M_\mathrm{core}$ is the
mass of the stellar core.

 During the common-envelope stage, underwent by a quarter to a half of
the systems, stars give their angular momentum very effectively to the
surrounding matter and spiral towards each other. The effectiveness of
the common-envelope stage is described by the parameter
$\alpha_\mathrm{CE}=\Delta E_\mathrm{b}/\Delta E_\mathrm{orb}$, where
$\Delta E_\mathrm{b}=E_\mathrm{grav}-E_\mathrm{thermal}$ is the binding
energy of the matter lost to the envelope and $\Delta E_\mathrm{orb}$ is
the decrease in the energy of gravitational interaction of the
approaching stars. The smaller parameter $\alpha_\mathrm{CE}$, the closer
components become after the CE stage. 

Further details on the wind scenarios and restrictions on the parameters
of the binary evolution can be found
in~\citet{lipunov_et1996b,lipunov_et1997,lipunov_et2005}. In the present
study, we take typical values of the parameters, also used in
\citet{bogomazov_et2007}.

To estimate the effective Kerr parameter at the start of collapse, we
need a relation between the radius and the mass of a
rotationally-synchronized core at the evolutionary stage, after which,
and until the collapse, the angular momentum is conserved. Following
previous studies, we believe that, at the He-burning stage, the star
rotation is fully synchronized with its orbital
motion~\citep[e.g.,][]{tutukov-cherep2003e,tutukov-cherep2004e,izzard_et2004,
podsiadlowski_et2004,vandenHeuvel-yoon2007,belczynski_et2007,zahn2008}.
We take into consideration that the carbon burning is likely to be
completed before the components of even a very close binary are
re-synchronized~\citep{masevich-tutukov1988book-e,zahn2008}. Thus,
during and after the C-burning stage, the period of the axial rotation
of the core is less than the orbital period. 

 One can estimate 
the radius of the core by the end of helium burning 
from the mass and temperature using the virial theorem:
\begin{equation}
\label{eq:mrco} R_\mathrm{c}=\frac{G\,\mu\, m_\mathrm{p}\,M_\mathrm{c}}{6\,k\,T},
\end{equation}
where $R_\mathrm{c}$ and $M_\mathrm{c}$ are the core radius and mass;
$T$ is the temperature of carbon burning, which is about $6\times
10^8$~K \citep{masevich-tutukov1988book-e}; $G$, the gravitational
constant; $\mu$, the average number of nucleons for a particle (equal to
15); $m_p$, the proton mass; $k$, the Boltzmann constant.  Here we use
the fact that CO-core of a massive star is not degenerate. 

The effective Kerr parameter is calculated following 
definition~\eqref{eq:kerr}. The moment of inertia can be written as
$I=k_I \,M_\mathrm{c}\,R_\mathrm{c}^2$, where a dimensionless parameter
$k_I$  likely lies within $\sim 0.1-0.4$ ($k_I=0.4$ for a uniform
spherical body; $k_I=0.1$ for polytrope 
spheres~\citep[][chap.~2.2]{zeld-blin-shakura}).

We perform population synthesis for different values of the evolutionary
parameters $\alpha_\mathrm{w}$, $\alpha_{CE}$, $\alpha_q$,  and
different wind models (Table~\ref{table:evol-param}). 
%%%%%%%%%%%%%%%%%%%%%%%%%%%%%%%%%%%%%%%%%%%%%%%%%%%%%%%%%%%%%%%%%%%%%%%%
\begin{table}
\centering
\caption{Parameters for scenarios of the population synthesis and the
resulting ratio $f$ of the number of the cores with $a_0>1$ 
to the total number of the cores.}
\label{table:evol-param}
\begin{tabular}{@{}ccccccc}
\hline
Population  & Wind model  & $\alpha_\mathrm{w}$ & $\alpha_\mathrm{CE}$ &
$\alpha_q$ & $k_I$  & $f$ \\ 
\hline
1   &  W1  &  0.3  &  0.5  &  0  &  0.4  &   0.10 \\
1a  &  W1  &  0.3  &  0.5  &  0  &  1.0  &    0.21 \\
1b  &  W1  &  0.3  &  0.5  &  0  &  0.1  &    0.02 \\
2   &  W1  &  0.1  &  0.5  &  0  &  0.4  &   0.10 \\
3   &  W1  &  0.7  &  0.5  &  0  &  0.4  &   0.09 \\
4   &  W1  &  0.3  &  0.5  &  2  &  0.4  &   0.05 \\
5   &  W1  &  0.3  &  1.0  &  0  &  0.4  &   0.10 \\
6   &  W2  &  ---  &  0.5  &  0  &  0.4  & 0.003 \\
\hline
\end{tabular}
\end{table}
%%%%%%%%%%%%%%%%%%%%%%%%%%%%%%%%%%%%%%%%%%%%%%%%%%%%%%%%%%%%%%%%%%%%%%%%
%%%%%%%%%%%%%%%%%%%%%%%%%%%%%%%%%%%%%%%%%%%%%%%%%%%%%%%%%%%%%%%%%%%%%%%%
\begin{figure}
  \begin{center}
   \resizebox{!}{0.35\textwidth}{\includegraphics{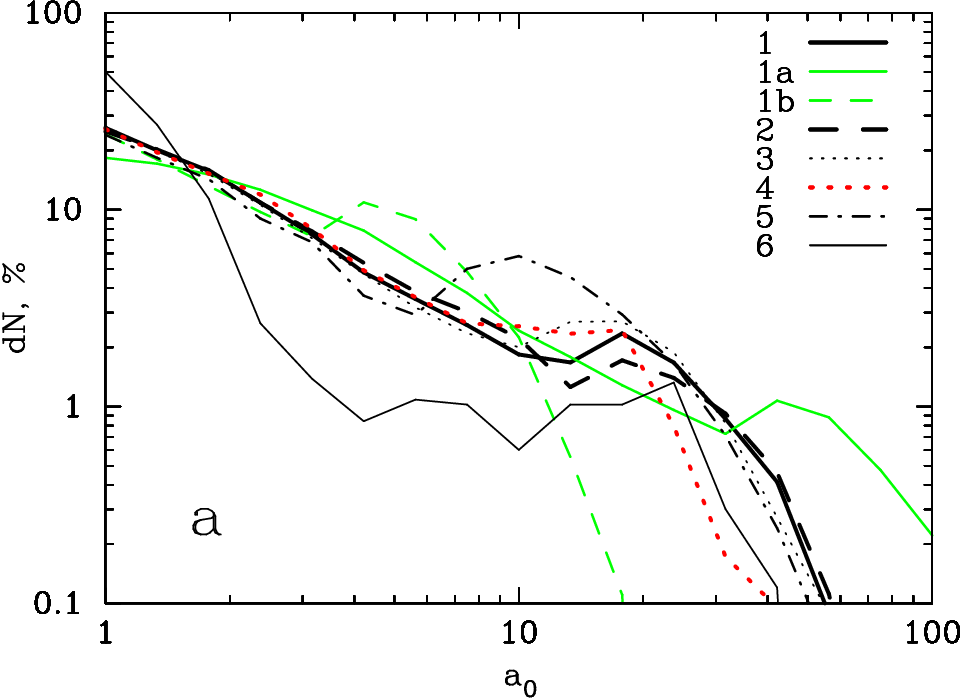}}
\resizebox{!}{0.35\textwidth}{\includegraphics{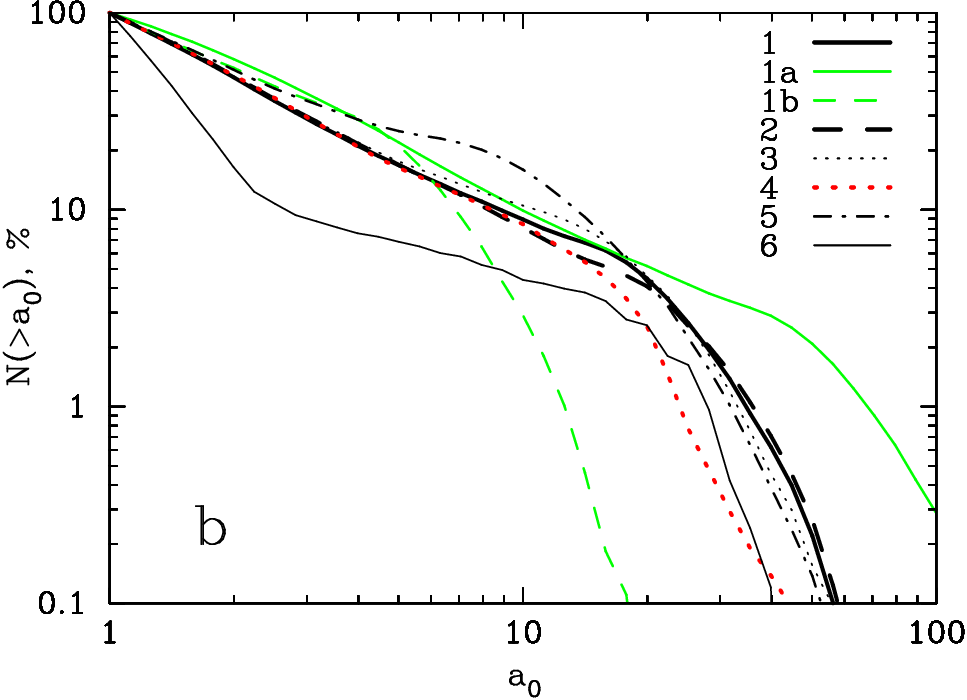}}
   % arguments: {xsize}{ysize}  % Insert PS-FRAG to see for more options
 \end{center} \caption{Probability density (a) and cumulative
distribution (b) of the effective Kerr parameter ($a_0>1$) of the CO-cores 
of WR stars 
in binary systems with different evolutionary parameters and $k_I$
listed in Table \ref{table:evol-param}.
Numbers on the vertical axis are normalized to the total
number of the cores with $a_0>1$, which is different for each
population.}
\label{fig:pop_result}
\end{figure}
%%%%%%%%%%%%%%%%%%%%%%%%%%%%%%%%%%%%%%%%%%%%%%%%%%%%%%%%%%%%%%%%%%%%%%%%
Fig.~\ref{fig:pop_result} presents the resulting distributions of the
CO-cores of WR-stars against the value of the effective Kerr
parameter. 

The last column in Table~\ref{table:evol-param} presents the
resulting ratio of the number of the cores with the effective Kerr
parameter greater than unity to the total number of the cores. Though
the absolute rate of GRBs is not the subject of our study, and it cannot
be inferred here, we note that the relative frequencies of GRB
production in different binary evolution scenarios are reflected by the
numbers in the last column of Table~\ref{table:evol-param}. An influence
of the possible magnetic coupling between the core and the slower
rotating envelope, which might brake core rotation up to several times
after the end of helium burning~\citep[see,
e.g.,][]{vandenHeuvel-yoon2007}, is illustrated by population `1b'. For
`1b', we decrease the coefficient of inertia $k_I$; effectively, this
describes some loss of the angular momentum of the core and leads to
lower number of fast-rotating cores. Note that our numerical estimates
agree with those of \citet{belczynski_et2007} who study the evolution of
\hbox{Population III} binaries\footnote{ See table 2 of their work,
model Mod08. This model is calculated with the zero magnetic coupling
between the core and the envelope, for the flat initial mass ratio
distribution. The fraction of the cores with the effective Kerr
parameter of the inner $7\,M_\odot$ greater than unity is about $0.1$ --
compare columns $\eta_1$ and $\eta_3$ -- just what we observe in our
Table~\ref{table:evol-param} for weak wind models `1', `2', `3', and `5'
with $k_I$ of the same order as in \citet{belczynski_et2007}, who  use
for the exposed He-cores $ k_I = 0.2-0.3$.}.

For the sets of parameters, used in populations `1', `4', and
`5',~\citet{bogomazov_et2007} performed population synthesis and
estimated absolute rates of collapse events with orbital periods less
than \hbox{$\sim 1$~day}. This corresponds to high values of the effective
Kerr parameter according with 
$$
a_0 \sim 0.3 \,\frac{M_\mathrm{c}}{M_\odot}\,
\frac{1\,\mbox{day}}{P_\mathrm{orb}}\, \frac{k_I}{0.4} \, ,
$$
which follows from equations \eqref{eq:kerr} and \eqref{eq:mrco}. Close
binaries containing a black hole, a main-sequence star, or a
nondegenerate star filling its Roche lobe contribute to this result with
different rates. The rates obtained by them agree with the observed
total GRB rate derived implying a solid angle correction for GRB
emission~\citep{podsiadlowski_et2004}.

\section{Calculation of spinar models}\label{s:spinar_calculation}

\subsection{A two-stage collapse of a spinar}

Here we describe the principles of the spinar paradigm~\citep[for more
details, see][]{lipunov-gorb2007,lipunov-gorb2008}. Consider the
collapse of a rotating object. When there is no sufficient internal
pressure to balance the gravitational force, the body will contract
until it encounters a centrifugal barrier at a radius $R_\mathrm{sp}$
estimated from the relation:
\begin{equation}
\omega^2 \,R_\mathrm{sp} = \frac{G\,M}{R_\mathrm{sp}^2}\, ,
\label{eq:centrif_bar}
\end{equation}
where $\omega$ is the angular velocity. At this point we say that a
spinar forms. Actual dynamical behavior of the configuration can be very
complex at this transition; however, the spinar model results in the
correct amount of the energy released in the process -- half of the
gravitational energy. From equation~\eqref{eq:kerr} one gets:
\begin{equation}
R_\mathrm{sp} = {a_0^2}\,\frac{G\,M}{c^2}\, .
\label{eq:Rsp}
\end{equation}
This radius is to be considered as a characteristic size in the
equatorial plane\footnote{In the spinar equations it is assumed that
$k_I=1$. This value ensures self-consistency of a calculated spinar
evolution when the spinar's size approaches $G\,M/c^2$.}. The effective
Kerr parameter is virtually constant during the first stage.

The evolution of a spinar proceeds as its angular momentum decreases
because of magnetic and viscous forces producing a braking torque. A
spinar can be characterized by an average magnetic field which is
represented by the magnetic dipole moment $\mu$. The braking torque on a
spinar can be expressed by a general formula:
\begin{equation}
\diff{(M R^2 \omega)}{t} = -k_\mathrm{t}\,\frac{\mu^2}{R^3}\, 
\label{eq:break_torque}
\end{equation}
\citep[see also][chapter 5]{lipunov1992}, where $R$ is the spinar radius
and $k_\mathrm{t}$ is a dimensionless parameter of order of unity. Such
an approach describes a maximally effective mechanism for spin-down.
Note that the spinar's rotation speeds up with decreasing spin.

The magnetic dipole moment $\mu$ can be expressed using parameter
$\alpha_{\mathrm m} = B_0^2 \, R_0^4/(6\, G\, M_0^2)$ -- the ratio of
the magnetic energy to the gravitational energy -- a measure of
magnetization introduced by \citet{lipunov-gorb2007}. The initial value
of magnetic dipole moment $\mu_0=B_0 \,R_0^3/2$ can be rewritten as 
$\sqrt { \alpha_{\mathrm m}\, \,G}\, M_0\, R_0$, incorporating constants
of order of unity into $\alpha_{\mathrm m}$. Here zero-subscripts
indicate initial values of the parameters, and $B$ is the dipolar
strength of the magnetic field, which approximates the actual field
configuration. In the approximation of magnetic flux conservation, the 
magnetic parameter $\alpha_{\mathrm m}$ remains constant.

As the spinar contracts and its rotation speeds up, the
energy-release rate increases, but as the spinar's size approaches the
size of a limiting Kerr black hole, the magnetic field squeezes up
against the surface, and the spinar's power fades away. Thus the second
peak is produced. To account for the vanishing magnetic field, we use
the following approximate expression~\citep{ginzburg-ozernoi1964e,
lipunova1997e,lipunov-gorb2008}:
\begin{multline}
\mu = \mu_0 \left(\frac{R_0}{R}\right)^2 
\frac{\xi (R_0/r_\mathrm{g},a_0)}{\xi(R/r_\mathrm{g},a)}\, ,\\
\xi(x,a) = \frac{x_\mathrm{min}(a)}{x} + 
\frac{x_\mathrm{min}^2(a)}{2\,x^2} + \ln
\left(1-\frac{x_\mathrm{min}(a)}{x}\right),
\end{multline}
where \hbox{$r_\mathrm{g}=G\,M/c^2$}
and
\hbox{$\xi(x,a)\approx - (x_\mathrm{min}/x)^3/3$} 
if \hbox{$x \gg x_\mathrm{min}$} and 
\hbox{$\xi(x,a)\rightarrow - \infty$} 
if \hbox{$x \rightarrow x_\mathrm{min}$}, 
with \hbox{$x_\mathrm{min}(a)=1+\sqrt{1-a}$} for $a<1$ and 
\hbox{$x_\mathrm{min}(a)=1$} for $a\geq 1$.

\citet{lipunov-gorb2008} obtain the characteristic time-scale for the
second stage, or the time scale of the angular momentum losses
(c.f.~\ref{eq:centrif_bar} and \ref{eq:break_torque}):
\begin{equation}
t_\mathrm{am} = \frac{G\,M}{c^3}\,
\frac{a_0^3}{2\,k_\mathrm{t}\,\alpha_\mathrm{m}}\, .
\label{eq:teor_sep}
\end{equation}

The complete set of equations~\citep[see][]{lipunov-gorb2008} includes
all the main relativistic effects near the gravitational radius: 
dynamics in the Kerr metric, magnetic field extinction, time dilation,
and gravitational redshift. The `spinar engine' posesses properties
commonly considered necessary to launch jets: rotation, magnetic field,
relativistic velocities and energies. These jets can be observed as
GRBs. The magneto-dynamical model of the spinar provides a main-scale
pattern of time evolution for a GRB, ignoring its short-scale
variability. We do not make any assumptions about the specific nature of
the jets, but we will imply for self-consistency that there is an
efficiency of processing the burst energy to the jet, which can be about
1 per cent (see also Sect.~\ref{ss:kpds}).

\subsection{Characteristics of spinar `light curves'}

\begin{figure}
  \begin{center}
\resizebox{!}{0.35\textwidth}{\includegraphics{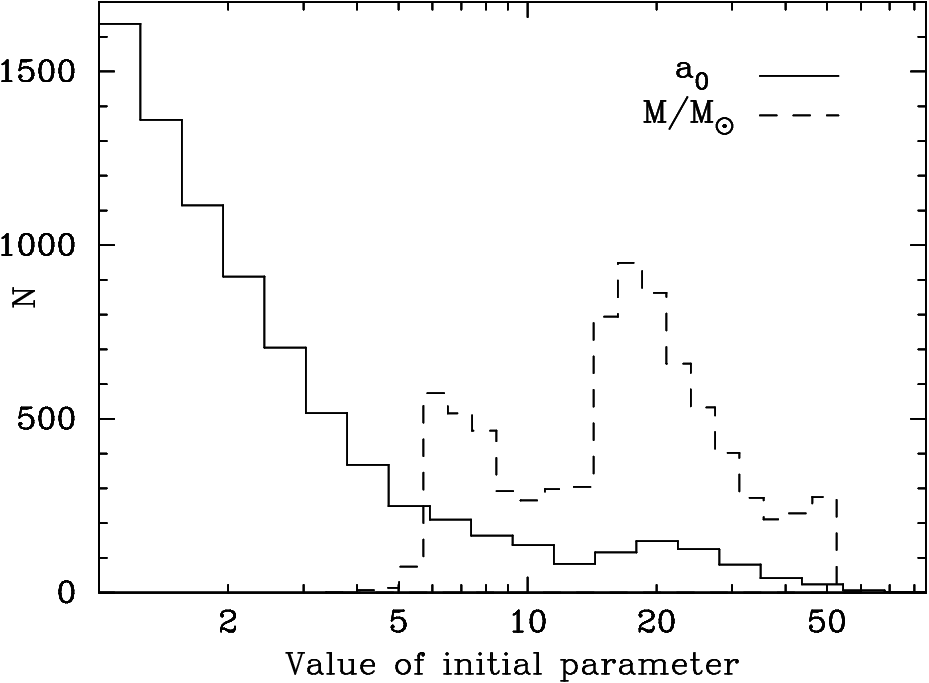}}
   % arguments: {xsize}{ysize}  % Insert PS-FRAG to see for more options
  \end{center}
  \caption{The distribution of the pre-collapse effective Kerr parameters (solid line)
and the distribution of the masses (dashed line) 
for the collapsing WR cores with $a_0>1$ for population
`1' (see Table~\ref{table:evol-param}). $N_\mathrm{total}=8000$.
}
  \label{fig:simulated_a_m}
\end{figure}

The input parameters to the model of collapse are the effective
dimensionless Kerr parameter of the core, the magnetic parameter, and
its mass. For example, Fig.~\ref{fig:simulated_a_m} shows the
distribution of the effective Kerr parameters and masses of the
cores of WR stars in close binary systems for population `1' (throughout
the paper we denote a result from population synthesis, i.e. a set of
massive cores, as a population with a corresponding number from the
first column of Table~\ref{table:evol-param}). Only cores with $a_0>1$
are counted in Fig.~\ref{fig:simulated_a_m}. We note that for
populations `1'-`5' the distribution of WR core masses with unrestricted
$a_0$ has one peak around $7-8$~M$_\odot$. More massive cores have an
advantage in acquiring large $a_0$ as they have a greater radius while
synchronized (see equation~\eqref{eq:mrco}); this results in the
multi-peak form of the mass distribution in
Fig.~\ref{fig:simulated_a_m}. At the same time, 
the mass distribution itself is not very important for the results of the
present work (it only shows itself in Fig.~\ref{fig:prec_distrib}a as a
three-branch disposition of points). For each pair $a_0$ and
$m=M/M_\odot$, we set a value of $\alpha_m$ randomly and uniformly
distributed on a logarithmic scale with limits $10^{-7}..10^{-2}$.
%%%%%%%%%%%%%%%%%%%%%%%%%%%%%%%%%%%%%%%%%%%%%%%%%%%%%%%%%%%%%%%%%%%%%%%%
%%%%%%%%%%%%%%%%%%%%%%%%%%%%%%%%%%%%%%%%%%%%%%%%%%%%%%%%%%%%%%%%%%%%%%%%

 To calculate spinar temporal evolution, we solve a system of
differential equations~\citep{lipunov-gorb2008} using a fourth-order
Rosenbrock method~\citep[][Chap.~16]{press_et2002} with tolerance
$10^{-5}$. We start integration of equations at radius $2\times 10^{3}\,
G\, M/c^2$. This value has little impact on the result, because the main
energy output of the first peak happens at the moment of spinar
formation, and the spinar radius does not depend on the initial radius
(see equation~\ref{eq:Rsp}). Simultaneously, while sampling the Scenario
Machine results, we set an upper limit on the effective Kerr parameter,
$a_0\approx 44$, to ensure that a core is not below the centrifugal
barrier at the beginning of a calculation. The lower limit on $a_0$ is
1. Coefficient $k_\mathrm{t}$ in the expression for the braking
torque~\eqref{eq:break_torque} is set to $1/3$.

%%%%%%%%%%%%%%%%%%%%%%%%%%%%%%%%%%%%%%%%%%%%%%%%%%%%%%%%%%%%%%%%%55
\begin{figure}
  \begin{center}
%  \rotatebox{-90}{%   do not forget ending % !!!
  \resizebox{!}{0.33\textwidth}{
\includegraphics[trim= 0cm 0cm 0cm 0cm, clip]{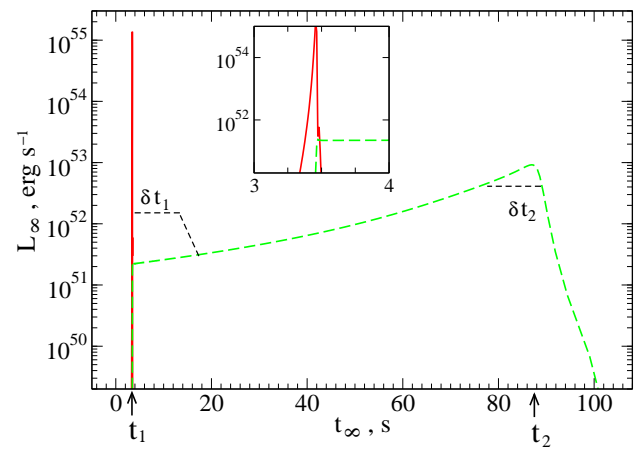}}
%}
  \end{center}
  \caption{
Energy release rate of a collapsing spinar (the sum of the solid and the
long-dashed line). Initial parameters are \hbox{$m=7$}, $a_0=5$, and
$\alpha_\mathrm{m}=10^{-4}$. Time and luminosity are in the frame of an
infinite observer. Left short-dashed profile represents characteristic
observed duration of the first peak. Right short-dashed horizontal line
is drawn at the 0.5 level from the peak value of $L_\infty$. The
overlaying graph zooms in a time interval near the first peak.}
  \label{fig:lc_pattern}
\end{figure}
%%%%%%%%%%%%%%%%%%%%%%%%%%%%%%%%%%%%%%%%%%%%%%%%%%%%%%%%%%%%%%%%%%%%%%%%%5
 Each trio of parameters, $a_0$, $\alpha_\mathrm{m}$, and $m$,  yields a
power curve with two pulses of different strength and duration. Various
patterns can be recognized, like plateaus, tails, one-peak patterns,
etc.

Consider as an example the burst evolution in Fig.~\ref{fig:lc_pattern}
that shows the rate of energy release as detected by an infinite
observer, corrected for the gravitational redshift and time dilation.
The solid line shows the dissipation rate of the kinetic energy at 
`impact' (i.e., the halt at the centrifugal barrier); the long-dashed
line shows the power released when the angular momentum is carried away.
One can see that the two stages are clearly separated by the relative
contribution of the two means of energy supply.

The burst pattern provides two useful temporal characteristics: the
separation between peaks $t_2-t_1$ and the duration of the second peak
$\delta t_2$. Generally, the computed duration of the first peak is very
short, because it corresponds to the time-scale of a dynamic model.
Thus, it should not be compared directly with a GRB jet duration. The
process of a jet development or propagation is not considered or
specified here. Duration of the second peak, estimated as shown in
Fig.~\ref{fig:lc_pattern}, might also underestimate the jet duration for
the same reason.

We assume that a model represents a GRB with precursor if it satisfies
the following condition: \hbox{$(E_1/\delta t_1) / L_2 <1$}. Here $E_1$
is the energy released during the first pulse, \hbox{$L_2
=L_\infty^\mathrm{max}$} -- the peak value for the second pulse. $E_1$
is obtained by integrating over the time interval during which
$L_\infty$ is greater than $l \times L_\infty^\mathrm{max}$, where $l$
is a fractional value. We usually set $l$ to 0.5 or 0.1.

A constraint on the actual duration of the first peak $\delta t_1$ comes
from observational data. It is widely accepted that the short and the
long GRBs have different underlying mechanisms or 
progenitors~\citep{norris_et2001,balazs_et2003,fox_et2005}. {\it Swift}
data indicate that long and short GRBs have different redshift
distributions with different median values of $z$: $0.4$ for short GRBs
and $2.4$ for long GRBs~\citep[][]{oshaughnessy_et2008,bagoly_et2006}.
We assume that both the first and the second peaks of the calculated
models contribute to the class of long GRBs. \citet{burlon_et2008} study
data for {\it Swift} GRBs with precursor with measured redshifts, all of
them having rest-frame precursor durations between 3 and 20 seconds.
Thus, we attribute factitious durations to the first peaks, generating
real values distributed randomly and uniformly in the above limits.

%%%%%%%%%%%%%%%%%%%%%%%%%%%%%%%%%%%%%%%%%%%%%%%%%%%%%%%%%%%%%%%%%%%%%%%%%%%5
\begin{figure}
 \begin{center}
\resizebox{!}{0.45\textwidth}{
\includegraphics[trim= 0cm 0cm 0cm 0cm,clip]{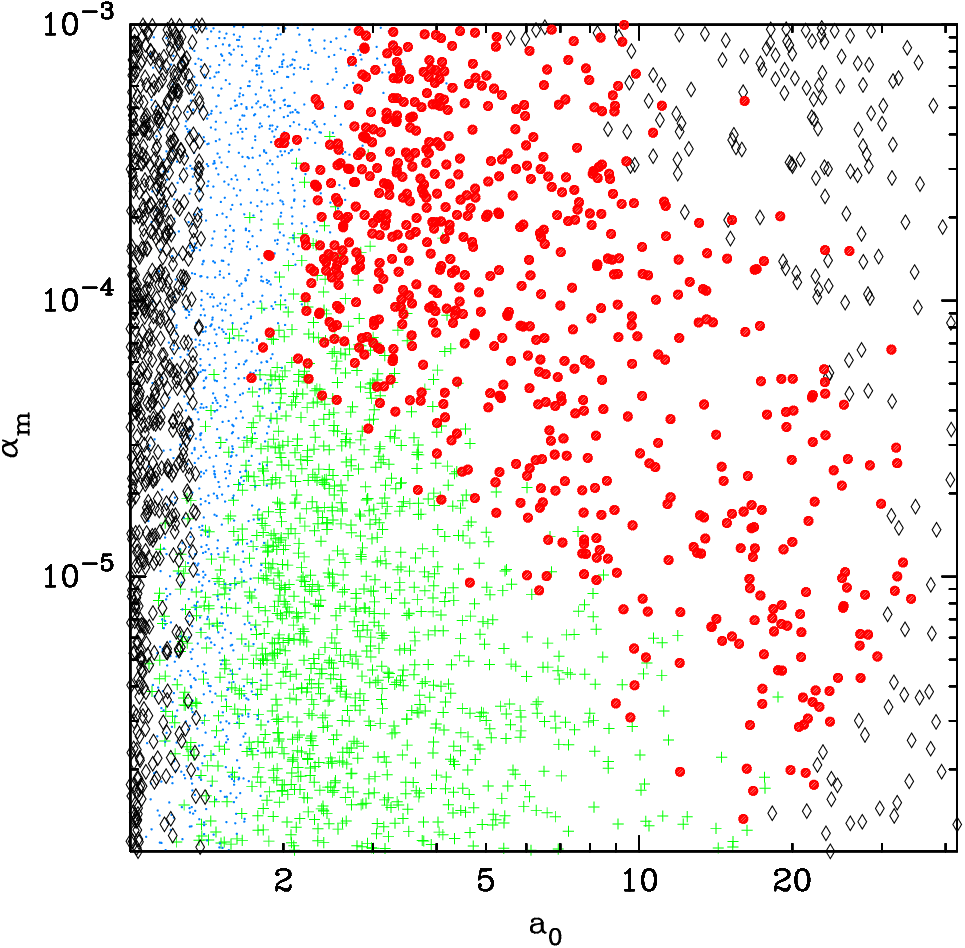}}
  \end{center}
  \caption{Magnetic parameter versus initial effective Kerr parameter  
for different classes of GRBs: with precursor (big red dots), with the
stronger first peak (green pluses), with single peak (black diamonds),
and merged peaks (small blue dots). 
Color is seen in the electronic version. For a comparison, the reader is
referred to the diagram in figure 2 of
~\citet{lipunov-gorb2008}.
}
  \label{fig:simulated_a_m_color}
\end{figure}
%%%%%%%%%%%%%%%%%%%%%%%%%%%%%%%%%%%%%%%%%%%%%%%%%%%%%%%%%%%%%

\section{Results}
\label{s:results}

For each population from Table~\ref{table:evol-param}, we calculate 8000
models within wide limits of $\alpha_\mathrm{m}$ ($10^{-7}..10^{-2}$).
Among the calculated models we can distinguish four classes of GRBs.
These are two classes with separate peaks: GRBs with precursor and GRBs
with a stronger first peak; third, GRBs with one pulse, produced by
models with $a_0$ either very large or close to unity; the last, GRBs
with very close pulses, meaning that the time separation between the
pulses is less than the maximum of two values: $0.1$~s (an arbitrary
chosen small value) and $(\delta t_1 + \delta t_2)/2$. 
Fig.~\ref{fig:simulated_a_m_color} shows distributions of the initial
model parameters for different classes of GRBs. Here precursors satisfy
the following condition: $1>(E_1/\delta t_1) / L_2 >0.01$. If a
precursor is very weak, then a GRB is re-classified and becomes a
single-peak GRB (diamonds in the top right corner of
Fig.~\ref{fig:simulated_a_m_color}).

%%%%%%%%%%%%%%%%%%%%%%%%%%%%%%%%%%%%%%%%%%%%%%%%%%%%%%%%%%%%%%%%%%%%%%%%%%%5
\begin{figure*}
  \begin{center}
   \resizebox{!}{0.9\textwidth}{
    \hskip -1cm
     \includegraphics[trim= 0cm 0cm 0cm 0cm,clip]{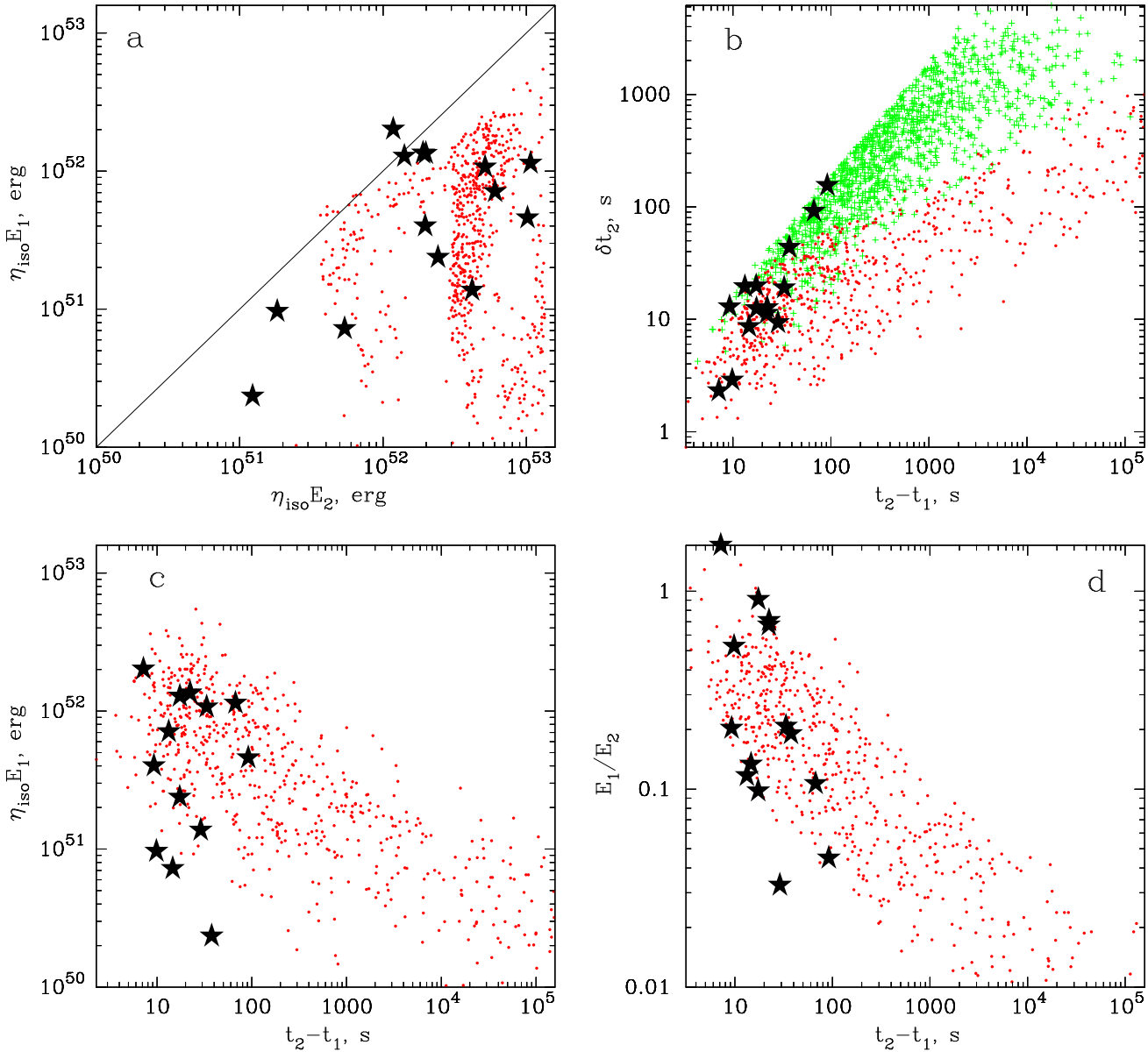}}
   % arguments: {xsize}{ysize}  % Insert PS-FRAG to see for more options
  \end{center}
  \caption{
(a) Energy of the first peak vs. energy of the second peak;
(b) duration of the second peak vs. separation between the peaks;
(c) energy of precursor vs. separation;
(d) precursor-to-main-peak energy ratio vs. separation.
Population `1' from Table~\ref{table:evol-param} is used.
Dimensionless coefficient 
$\eta_\mathrm{iso}=0.01$ for both peaks. Magnetic
parameter $10^{-6}<\alpha_\mathrm{m}<10^{-3}$.
Pluses (green in the electronic version) in (b) represent results for the
models with a stronger first peak. Stars are plotted for the data
taken from \citet{burlon_et2008}: rest-frame
magnitudes $E_\mathrm{iso}$, mostly in $15-150$~keV,  
separation times, calculated as 
($T_{1,\mathrm{main}} - T_{1,\mathrm{prec}})/(1+z)$, and 
durations of the main peaks  
($T_{2,\mathrm{main}} - T_{1,\mathrm{main}})/(1+z)$.
}
\label{fig:prec_distrib}
\end{figure*}
%%%%%%%%%%%%%%%%%%%%%%%%%%%%%%%%%%%%%%%%%%%%%%%%%%%%%%%%%%%%%%%%%%%%%%%%%%%%%5

In Fig.~\ref{fig:prec_distrib} we present characteristic time and energy
distributions for a class of models identified as GRBs with precursor.
The initial parameters for the models in Fig.~\ref{fig:prec_distrib} are
taken from population `1'~(see Sect.~\ref{s:popsyn}). The energies and
durations of the peaks are calculated for a level of $0.5$ of the peak
luminosity~(for illustration see Fig.~\ref{fig:lc_pattern}). In
Fig.~\ref{fig:prec_distrib} the energy of pulses is given as
$\eta_\mathrm{iso}\,E_\mathrm{P}\equiv E_\mathrm{iso}$, where
$E_\mathrm{P}$ is the calculated energy of a peak, P\,$=1$ or $2$, and
$E_\mathrm{iso}$ is the rest-frame isotropic bolometric energy, which
can be derived from values observed in a spectral energy band. Here we
set $\eta_\mathrm{iso}=0.01$ for both peaks in order to cover with the
modeled points the area that is occupied by the observed values adopted
from \citet{burlon_et2008} (stars in Fig.~\ref{fig:prec_distrib}).
Coefficient $\eta_\mathrm{iso}$ incorporates our freedom in choosing
level $l$ as well as the actual efficiency of the gravitational
collapse, and the fact that there are two jets.

\citet{burlon_et2008} study a sample of GRBs with precursor activity
detected by {\it Swift} and with known redshifts. All GRBs with
precursors belong to a class of long GRBs, as also found for the {\it
BATSE} catalogue by~\citet{koshut_et1995}. We use the data from Table 1
of~\citet{burlon_et2008} except for GRB~070306 that has two precursors
(see Sect.~\ref{ss:multi-precursors}). The observed time intervals are
recalculated to the rest-frame using known redshifts. One can see that
the separation times $t_2-t_1$ between the peaks calculated in the
spinar model reproduce very well the observed time lags and that longer
lags are also produced. From the practical point of view it is worth
noting that separation times calculated numerically agree with values
$t_\mathrm{am}$ defined by \eqref{eq:teor_sep}
quite closely, within an accuracy of 10\% for most of the models.
Formula~\eqref{eq:teor_sep} can be rewritten as 
\begin{equation}
t_{am} \approx 20 \, \frac{M}{10\, M_\odot}\,
\left(\frac{a_0}{3}\right)^3\,\left(\frac{\alpha_\mathrm{m}}{10^{-4}}\right)^{-1}\,
\left(\frac{k_\mathrm{t}}{1/3}\right)^{-1} \, \mathrm{s}.
\label{eq:teor_sep_s}
\end{equation}

In Fig.~\ref{fig:prec_distrib}b, we also show by pluses results for the
models with a stronger first peak. Some of these models can manifest
themselves as events with a precursor due to some dispersion of
$\eta_\mathrm{iso}$ and differences in the efficiencies of the two
peaks. The scatter of the observed values (stars in
Fig.~\ref{fig:prec_distrib}) can be also caused by different
efficiencies in the sources. In the context of
Fig.~\ref{fig:prec_distrib}b, we notice that a correlation with slope
about $1$ between emission duration time and previous quiescent time in
the observer frame was found by \citet{ramirez-ruiz-merloni2001} (see
their figure 3b) who investigated long and bright {\it BATSE} bursts
containing at least one quiescent interval in their time history (they
did not sample precursor events). Their slope roughly coincides with the
overal slope of the star-distribution in Fig.~\ref{fig:prec_distrib}b.

%%%%%%%%%%%%%%%%%%%%%%%%%%%%%%%%%%%%%%%%%%%%%%%%%%%%%%%%%%%%%%%%%%%%%%%%
\begin{table}
\centering
\caption{Relative frequency of GRBs with precursor for different
ranges of the modeled separation time between the precursor and the
main peak.
}
\label{table:prec_wind_models}
\begin{tabular}{cccc}
\hline
Population   &  \multicolumn{3}{c}{$t_2-t_1$,~s} \\
             &   $<100$ & $100-10^3$ & $100-10^6$ \\
\hline
1   & 0.08 &  0.04 & 0.10\\
1a  & 0.11 &  0.07 & 0.12\\
1b  & 0.08 &  0.09 & 0.14\\
2   & 0.09 &  0.05 & 0.10\\
3   & 0.08 &  0.05 & 0.12\\
4   & 0.09 &  0.05 & 0.11\\
5   & 0.07 &  0.05 & 0.17\\
6   & 0.02 &  $<$0.01& 0.03 \\
\hline
\end{tabular}
\end{table}

%%%%%%%%%%%%%%%%%%%%%%%%%%%%%%%%%%%%%%%%%%%%%%%%%%%%%%%%%%%%%%%%%%%%%%%%
In Table~\ref{table:prec_wind_models} we give the approximate ratios of
the number of GRBs with precursor to the number of all GRBs for
different binary evolution parameters. The fraction of the models with
precursor of separation $<100$~s is $8-10$~per cent for most of the
populations. All GRBs include two-peak GRBs, single peak GRBs, and GRBs
with very close peaks. To calculate these ratios, only precursors that
are `strong enough to be detected' are counted; that is, they satisfy an
arbitrary chosen condition: $(E_1/\delta t_1) / L_2 >0.01$.

The numbers in the Table~\ref{table:prec_wind_models} should be regarded
as rough estimates, as they are based on an assumption that a modeled
power curve (as, for example, in Fig.~\ref{fig:lc_pattern}) provides a
realistic relation between $L_1$ and $L_2$. If we halve the efficiency
$\eta_\mathrm{iso}$ of the precursor, the relative number of precursors
becomes greater but not significantly ($\sim 10$~per cent for population
`1', for example). To provide some glimpse of the significance of the
numbers, we can say that, if we change the level for calculating peak
duration and energy to $l=0.1$ and keep other parameters unchanged, then
the number of GRBs with precursor drop to 6 per cent for population `1'.

\begin{figure}
  \begin{center}
   \resizebox{!}{0.35\textwidth}{
    \includegraphics[trim= 0cm 0cm 0cm 0cm,clip]{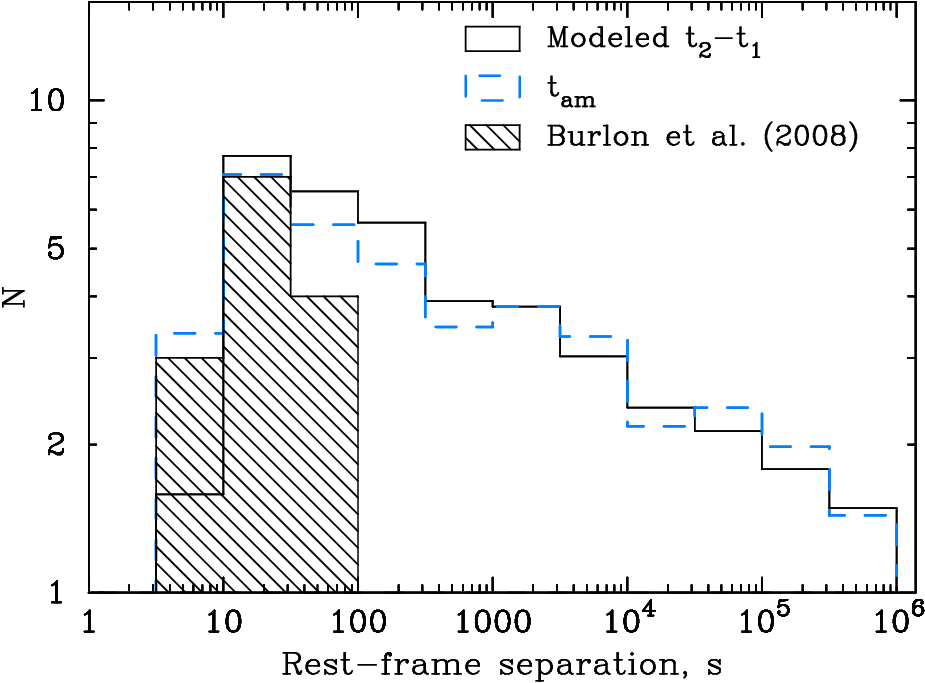}}
   \end{center}
  \caption{Distribution of the separation times between the precursor
and the main pulse.
Empty histograms show model estimates and are vertically scaled with the
same area under the solid and the dashed line. Hatched histogram represents
rest-frame separation times ($T_{1,\mathrm{main}} - 
T_{1,\mathrm{prec}})/(1+z)$ from \citet{burlon_et2008}. The set of
models is the same as in Fig.~\ref{fig:prec_distrib}.}
\label{fig:sep}
\end{figure}

\begin{figure}
  \begin{center}
   \resizebox{!}{0.35\textwidth}{
    \includegraphics[trim= 0cm 0cm 0cm 0cm,clip]{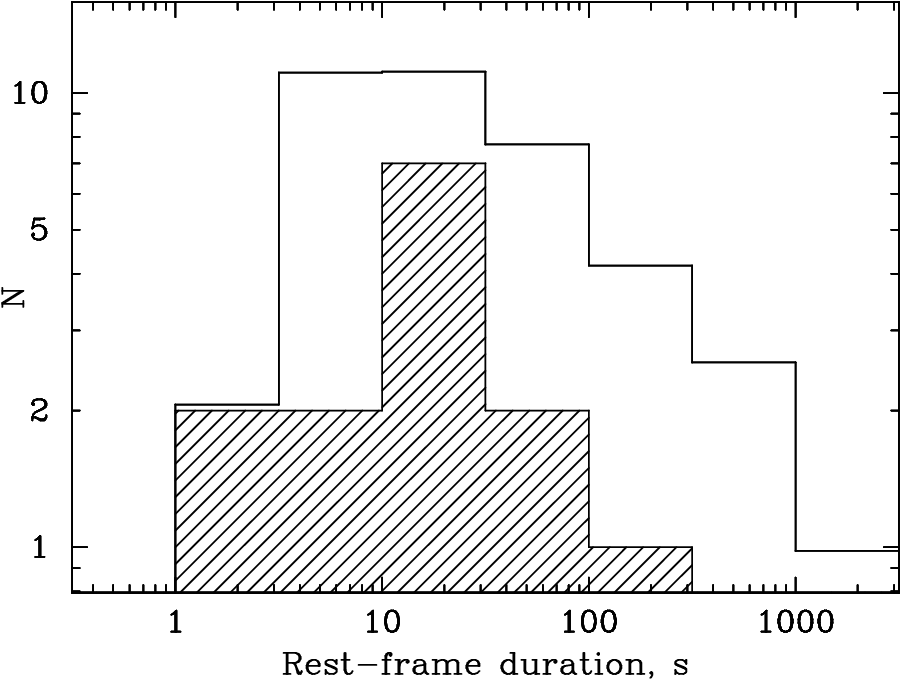}}
   % arguments: {xsize}{ysize}  % Insert PS-FRAG to see for more options
  \end{center}
  \caption{Distribution of the duration of the main pulse.
Empty histogram shows model results for events with precursors and is
vertically scaled.
Hatched histogram represents rest-frame duration times
$(T_{2,\mathrm{main}}-T_{1,\mathrm{main}})/(1+z)$
taken from \citet{burlon_et2008}. The set of
models is the same as in Fig.~\ref{fig:prec_distrib}.}
\label{fig:dur}
\end{figure}

 Normalized histograms for distributions of modeled separation times and
main (i.e., second) peak durations are plotted in Figs.~\ref{fig:sep}
and~\ref{fig:dur}. Observational data from \citet{burlon_et2008} is also
presented. The modeled duration of the second peak might be a minimum 
estimate of the actual duration because no radiation and
magneto-hydrodynamic effects are incorporated in the present model of a
GRB.

%%%%%%%%%%%%%%%%%%%%%%%%%%%%%%%%%%%%%%%%%%%%%%%%%%%%%%%%%%%%%%%%%%%%%%%%%%%5
%%%%%%%%%%%%%%%%%%%%%%%%%%%%%%%%%%%%%%%%%%%%%%%%%%%%%%%%%%%%%%%%%%%%%%%%%%%%%5

%%%%%%%%%%%%%%%%%%%%%%%%%%%%%%%%%%%%%%%%%%%%%%%%%%%%%%%%%%%%%%%%%%%%%%%%%%%5
\begin{figure}
  \begin{center}
   \resizebox{!}{0.45\textwidth}{
    \includegraphics[trim= 0cm 0cm 0cm 0cm,clip]{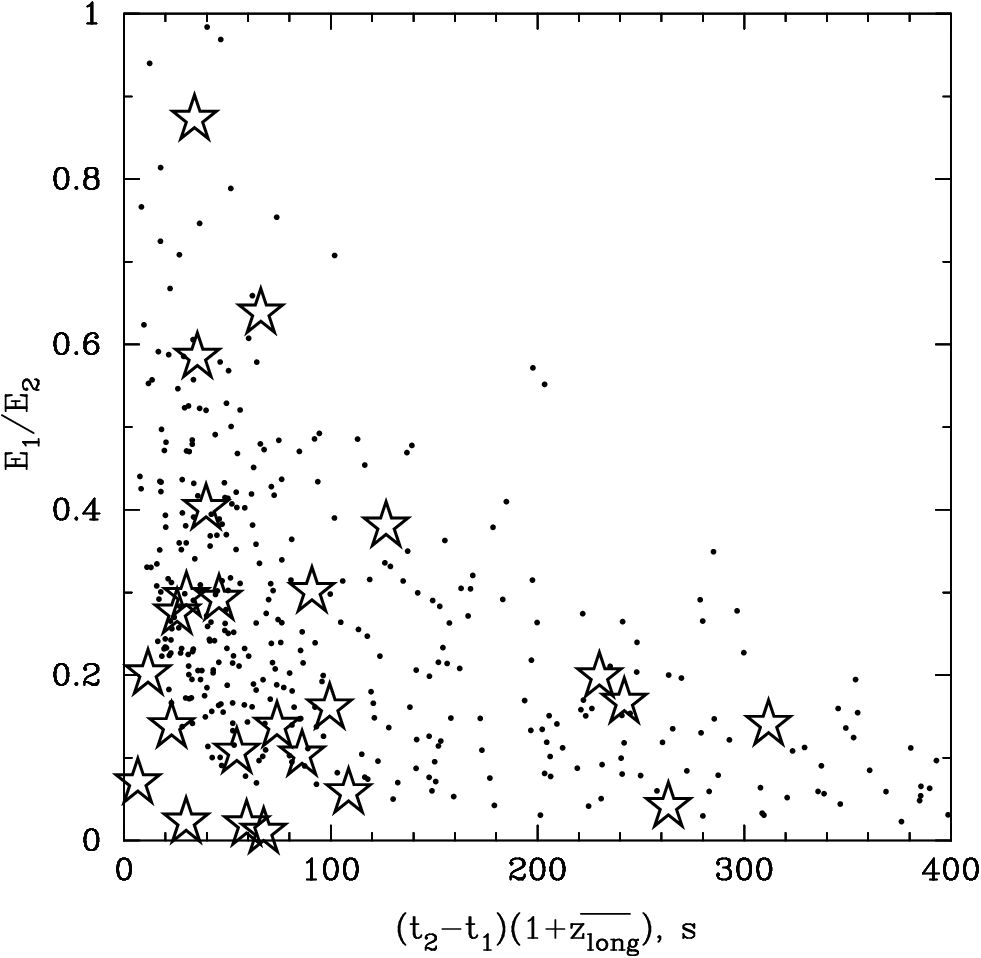}}
   % arguments: {xsize}{ysize}  % Insert PS-FRAG to see for more options
  \end{center}
  \caption{Ratio of the energy released during the precursor and the main
peak versus separation time multiplied by the constant factor. 
Dots represent model results for
GRBs with precursor for population '1'. 
The average value of the redshift for long GRBs
 $\overline{z_\mathrm{long}}=2.4$. Stars are  adopted from 
figure 13 of~\citet{koshut_et1995} that shows the 
ratio of the precursor total counts to the main episode total
counts versus separation time
between detectable emission for {\em BATSE} GRBs with precursor.}
\label{fig:E12}
\end{figure}

%%%%%%%%%%%%%%%%%%%%%%%%%%%%%%%%%%%%%%%%%%%%%%%%%%%%%%%%%%%%%%%%%%%%%%%%%%%%%5
The longest observed separation between the peaks of a precursor and a
main pulse, $\sim 400$~s, was reported by~\citet{koshut_et1995} but no
conclusion about the rest-frame time separation can be made without
knowing $z$. Nevertheless, our results are in apparent agreement with
the data from figure 13 of~\citet{koshut_et1995} that shows the ratio of
the precursor total counts to the main episode total counts versus
separation time between detectable emission. The distribution in
Fig.~\ref{fig:E12} is constructed for population `1' for analogous
values where we try to emulate cosmologically dilated separation times
by multiplying all values $(t_2-t_1)$ by a constant factor.

%%%%%%%%%%%%%%%%%%%%%%%%%%%%%%%%%%%%%%%%%%%%%%%%%%%%%%%%%%%%%%%%%%%%%%%%%%%5
%%%%%%%%%%%%%%%%%%%%%%%%%%%%%%%%%%%%%%%%%%%%%%%%%%%%%%%%%%%%%%%%%%%%%%%%%%%%%5

Summarizing, the spinar paradigm explains observed precursor separation
times and predicts that there are very early precursors with rest-frame
time separations more than 100 s, which is the maximum determined
rest-frame separation so far~\citep{burlon_et2008}. Our results show
that the rate of GRBs with precursor, which have separations longer by
up to one order of magnitude than those observed so far, is about half
the detected rate~(compare the second and third columns of
Table~\ref{table:prec_wind_models}). The number of modeled GRBs with
precursor with lead times in the range from $100$ s to $10^6$~s is
comparable to the frequency of precursor events observed so far. At the
same time, the expected average energy output of a precursor drops with
increasing separation time as illustrated by
Fig.~\ref{fig:prec_distrib}c.

One should keep in mind that a rest-frame time interval as, for example,
$100$~s, is dilated to $(1+z)\times 100$~s in the observer's frame. At
this point, an important general conclusion can be made: GRBs at high
redshifts will show precursor events which should arrive $(1+z)\times
100 \sim 1000$~s in advance of the main event. This prediction is very
strong as it does not depend on a GRB model.

\section{Discussion}
\label{s:discussion}

The relative frequencies of the GRBs with precursor, obtained in the
spinar paradigm, agree reasonably well with observational facts.
\citet{koshut_et1995} find precursors in $\sim 3$~per cent of {\it
BATSE} long and short GRBs detected before May 1994; \citet{lazzati2005}
finds that precursors are present in $\sim 20$~per cent of bright, long
GRBs in the final {\it BATSE} catalog; \citet{burlon_et2008} obtains
that about $14$~per cent from 105 GRBs with measured redshift, observed
by {\it Swift} before March 2008, have precursor activity. Note that
these researches use different criteria for precursor search and
selection. All these data agree quite well with the modeled ratios
within an accuracy of $5-10$ per cent~(see
Table~\ref{table:prec_wind_models}). Note that there is no significant
difference in the ratios between the different binary evolution
scenarios (for wind model W1) and different $k_I$.
Apparently, the strong wind scenario experiences the
most difficulty in describing the reported ratios (population `6').

To speak of a more robust result, the spinar model provides a clear
explanation of the diversity of observed time separtions between
precursor and main peak, which, moreover, can be related to the main
physical parameters of a collapsing object in a simple way
(equation~\ref{eq:teor_sep} or~\ref{eq:teor_sep_s}). To our knowledge, explaining both 
relatively short ($\sim 10$~s) and long ($\sim 100$~s) switch-off times
presented difficulties for most of the GRB engine models~\citep[see][and
references therein]{wang-meszaros2007,drago-pagliara2007}.

Accretion has so far been neglected by us, but could be easily
incorporated in the equations, as pointed out
by~\citet{lipunov-gorb2008}; this is a topic for a forthcoming paper. 
As a preliminary result, we found that accretion of up to $\gtrsim
1~$M$_\odot$ during the overall GRB time ($10-100$~s) does not change
qualitatively any of the results presented.

\subsection{Multi-precursors}
\label{ss:multi-precursors} Among 15 GRBs with precursor studied
by~\citet{burlon_et2008} there is one (GRB~070306) with two precursors.
The first precursor is $\sim 105$~s from the second precursor, which is
ahead of the main pulse by $\sim 95$~s. Each pulse is stronger than the
previous one. This GRB is not shown in the figures in the present paper.

 One way to explain such a phenomenon is an unstable regime of jet with
a constantly operating central engine. A mechanism of a modulated
relativistic wind, resulting in long time gaps in emission, was
considered by~\citet{ramirez-ruiz_et2001}. 

Other possibilities appear if we consider the basic elements of the
spinar paradigm. As we argue in the Introduction, the course of a
collapse is affected by the initial distribution of the angular momentum
in the rotating object. In this regard, spinar and collapsar models are
`opposite' options. In reality there is most probably an assortment of
mixed scenarios. One possibility, provided by a particular angular
momentum distribution in a pre-collapse core, is that a black hole forms
first (primary pulse), then a spinar (heavy accretion disc) forms around
it (second pulse), followed by a fatal collapse (hyper-accretion) of a
spinar into a black hole (third pulse). Another explanation for a
three-stage collapse can be that an accretion disc aggregates at some
stage of a spinar evolution. Here we should mention the `initial jet and
fallback collapsar' scenario~\citep{wang-meszaros2007} with its
characteristic disc time of $\sim 100$~s. In its context, we can
speculate why the majority of GRBs with precursor (or GRBs with one or
more quiescent periods, see~\citet{ramirez-ruiz-merloni2001}) are not
three-step events. The fallback scenario works for explosions that are
not too weak or too strong according to \citet{fryer1999}. An explosion
which is too weak leads to a direct collapse to a BH, or a spinar if it
rotates fast enough. Rotating progenitors possibly favor `direct
collapse to a spinar' without a huge explosion; this view is supported
by general and computational~\citep{monchmeyer1991, yamada-sato1994}
arguments that rotation weakens the bounce and hence the explosion.

Numerical studies yield further options for gravitational collapse: 
fragmentation of a rotating collapsing body, a multiple black hole
system formation~\citep[see, for
example,][]{berezinskii_et1988,imshennik1992e,zink_et2007}, and its
eventual merger. Such scenarios \citep[see also][]{king_et2005} can
provide a burst with several pulses, but are beyond the scope of the
present work.

\subsection{Lower limit on a precursor energy}
\label{ss:kpds} Let us provide some relations for the energetics of the
precursor jet. The precursor jet is apparently the first to break
through a shell that probably surrounds the central object.
\citet{lipunov-gorb2008} arrive at a condition for a jet to penetrate
the surrounding shell. The jet breaks through if the momentum imparted
on a part of the shell is greater than the momentum corresponding to the
escape velocity:
\begin{equation}
\beta \,E_\mathrm{P} = E_\mathrm{jet}^\mathrm{p} >
\frac{\Omega_\mathrm{s}}{4\pi} \,M_\mathrm{shell} \, \sqrt
\frac{2GM_\mathrm{core}}{R_\mathrm{shell}} \, c\, ,
\label{eq:push_jet}
\end{equation} where $\beta$ is the efficiency of processing the energy
of the collapse to the pushing jet, $c$ is the speed of the jet, which
is approximately the speed of light, ${\Omega_\mathrm{s}}/{4\pi}$ is the
portion of the shell's surface subject to the jet. Substituting
universal constants, one has:
\begin{equation}
E_\mathrm{jet}^\mathrm{p} \gtrsim 6\times 10^{50} \,
\frac{{\Omega_\mathrm{s}}}{0.01\times 4\pi} \, m_\mathrm{shell}\,
\left(\frac{R_\mathrm{shell}}{2\times10^{3}\,r_\mathrm{g}}\right)^{-1/2}\,
\mathrm{erg},
\end{equation}
where $m_\mathrm{shell}$ is in solar masses. 

Now consider the relations for the observed jet:
\begin{equation}
E_\mathrm{jet}^\mathrm{ob} = k_\gamma \, E_\mathrm{jet}^\mathrm{p},
\quad
E_\mathrm{jet}^\mathrm{ob}   \equiv E_\mathrm{iso} \,
\frac{\Omega_\mathrm{ob}}{4\pi} \, .
\label{eq:ob_jet}
\end{equation}
It is natural to assume that the energy of the jet punching the shell is
not the same as that of the observed $\gamma$-jet and to introduce a
factor $k_\gamma<1$ between them, which depends, among other things, on
the spectral band observed. Here $\Omega_\mathrm{ob}$ is the solid angle
corresponding to the observed jet's opening angle, $E_\mathrm{iso}$ is
the rest-frame isotropic bolometric energy, which can be derived from
the values observed in a spectral energy band.

From the above we derive the minimum value $E_\mathrm{iso}$ that corresponds to a jet
still capable of going through the shell:
\begin{equation}
E_\mathrm{iso} (min) =  6\times 10^{52} k_\gamma\,
\frac{\Omega_\mathrm{s}}{\Omega_\mathrm{ob}} \,   m_\mathrm{shell}\,
\left(\frac{R_\mathrm{shell}}{2\times10^{3}\,r_\mathrm{g}}\right)^{-1/2}\,\mathrm{erg}.
\label{eq:Eiso_min}
\end{equation}

Let us now illustrate this with some numbers. 
Fig.~\ref{fig:prec_distrib}a possibly indicates that $E_\mathrm{iso}
(min) \sim 10^{50}$~erg from the data of \citet{burlon_et2008}.
Comparing with equation~\eqref{eq:Eiso_min}, one can obtain an implicit
estimate of $k_\gamma$ for the observational data mentioned as low as
$0.1- 1$~per cent, if $\Omega_\mathrm{s} \sim \Omega_\mathrm{ob}$ and
$m_\mathrm{shell}\sim 1$.

In Sect.~\ref{s:results}, we introduced the parameter
$\eta_\mathrm{iso}$ to match the modeled energy of the pulses with the
rest-frame isotropical values $E_\mathrm{iso}$. We estimated it as
\hbox{$\eta_\mathrm{iso}\sim 0.01$}. Using \eqref{eq:push_jet} and
\eqref{eq:ob_jet}, $\eta_\mathrm{iso}$ can be expressed as follows:
\begin{equation}
\eta_\mathrm{iso} \equiv \frac{E_\mathrm{iso}}{E_\mathrm{P}} =
\frac{4\pi}{\Omega_\mathrm{ob}}\, k_\gamma\, \beta\, .
\label{eq:kpd}
\end{equation} If the half-opening angle of a jet $\sim
0.1$~rad~\citep[e.g.,][]{meszaros2006} and $k_\gamma \sim 0.001-0.01$ 
then $\Omega_\mathrm{ob} ~\sim 0.03~\mathrm{sr}\sim 4\,\pi/400$ and we
come at an estimate for the effieciency of processing the energy of the
collapse to the jet in the shell: $\beta\sim\eta_\mathrm{iso}\sim
0.01$.

\section{Conclusion}

Currently, much effort is put into numerical computations of GRBs and 
simulations of jets emerging from rotating magnetized configurations, to
name just a few, by \citet{lyutikov2006,barkov-komissarov2008,
mckinney-blandford2009,takiwaki_et2009} and others. We believe that the
progress with the spinar mechanism for GRBs demands a further
development by means of a MHD simulation.

In the present work, we calculate many models of two-stage spinar
evolution provided that fast rotating stellar cores are produced in
binary systems due to a tidal synchronization mechanism. The
pre-collapse parameters of WR stars are calculated by the Scenario
Machine dedicated to the population synthesis of binary
systems~\citep{lipunov_et1996a, lipunov_et2007}.

We analyse the resulting set of GRBs, which can be classified into
different classes. For GRBs with precursor, separation times between the
precursor and the main pulse are fully consistent with observational
data. We find that precursor events with the rest-frame time separation from
the main pulse $<100$~s occur in about $10$ per cent of the modeled GRB
events, for the weak wind scenarios of binary evolution and
independently of the poorly established value of $k_I$. This rate,
however, depends on the specifics of the jet generation and its
reliability is limited by the framework of our simple model. Taking this
into account, we think the modeled rate agrees very well with the
observed values. Comparison of the time and energetic characteristics of
the modeled bursts with the observational data for GRBs with measured
redshifts~\citep{burlon_et2008} is made and a good general agreement is
found.

We predict that very early precursors with a rest-frame time separation
more than 100~s should exist, and the number of GRBs with precursor with
separations in the range from $100$ s to $1000$~s is about half the
number of those observed so far. Super-early primary pulses, up to 
$10^{6}$~s, are also found in the model; they, however, may be too weak
to be detected or to develop as a jet.

Finally, whether or not the present model works in all cases, the
observational evidence is that rest-frame separation times of $\sim
100$~s take place. High-redshift long GRBs, up to $z \gtrsim 10$, are
expected to be detected by future
experiments~\citep[e.g.,][]{salvaterra_et2008}.  Thus, we
confidently expect precursor events arriving up to $\gtrsim$1000~seconds
in advance of the main GRB episode.

\section*{Acknowledgments}
We thank the anonymous referee for the useful suggestions
which have helped to improve the manuscript.
 We are grateful to A. Tutukov and R. Porcas for helpful comments. GVL
is partly supported by the Russian Foundation for Basic Research
(project 09-02-00032). AIB is supported by the State Program of Support
for Leading Scientific Schools of the Russian Federation (grant
NSh-1685.2008.2) and the Analytical Departmental Targeted Program `The
Development of Higher Education Science Potential' (grant
RNP-2.1.1.2906). GVL is grateful to Y. Kovalev for support and the
Offene Ganztagesschule of the Paul-Klee-Grundschule (Bonn, Germany) and
Stadt Bonn for providing a possibility for her full-day scientific
activity.

\bibliographystyle{mn2e}

%\bibliography{lipunova}

%\end{document}

\end{document}